\documentclass[aps,prl,twocolumn,showpacs,amsmath,amssymb,superscriptaddress,reprint]{revtex4-1}

\usepackage[utf8]{inputenc}
\usepackage{graphicx}
\usepackage[colorlinks=true,citecolor=blue]{hyperref}
\usepackage{units}
\usepackage{verbatim} %mehrzeilige Kommentare

\usepackage{pdfpages}
\usepackage{pgffor}

\renewcommand{\vec}[1]{\mathbf{#1}}
\newcommand{\kBT}{k_\text{B}T}

\def\up{\uparrow}
\def\down{\downarrow}
\def\kB {k_\text{B}}

\begin{document}
\title{Fingerprint of topological Andreev bound states in phase-dependent heat transport}
\author{Bj\"orn Sothmann}
\affiliation{Institute for Theoretical Physics and Astrophysics, University of Würzburg, Am Hubland, 97074 Würzburg, Germany}
\author{Ewelina M. Hankiewicz}
\affiliation{Institute for Theoretical Physics and Astrophysics, University of Würzburg, Am Hubland, 97074 Würzburg, Germany}

\date{\today}

\begin{abstract}
We demonstrate that phase-dependent heat currents through superconductor-topological insulator Josephson junctions provide a useful tool to probe the existence of topological Andreev bound states, even for multi-channel surface states.
We predict that in the tunneling regime topological Andreev bound states lead to a minimum of the thermal conductance for a phase difference $\phi=\pi$, in clear contrast to a maximum of the thermal conductance at $\phi=\pi$ that occurs for trivial Andreev bound states in superconductor-normal metal tunnel junctions.
This opens up the possibility that phase-dependent heat transport can distinguish between topologically trivial and nontrivial 4$\pi$ modes.
Furthermore, we propose a superconducting quantum interference device geometry where phase-dependent heat currents can be measured using available experimental technology.
\end{abstract}

%\keywords{}
\maketitle

\paragraph{Introduction.--}
At the interface between a normal metal and a superconductor, an electron with an energy inside the superconducting gap gets reflected from the superconductor as a hole in a process called Andreev reflection~\cite{andreev_thermal_1964}.
In a superconductor-normal metal-superconductor (S-N-S) junction, this electron-hole conversion leads to the formation of Andreev bound states with discrete energies~\cite{andreev_electron_1966}.
In the case of a short, one-dimensional junction there is exactly one pair of such bound states with energy $\varepsilon=\pm\Delta\sqrt{1-D\sin^2\phi/2}$ where $D$ denotes the transmission probability of the junction in the normal state, $\Delta$ is the absolute value of the superconducting pair potential and $\phi$ the phase difference across the junction~\cite{beenakker_universal_1991}.

Recently, there has been a growing interest in Josephson junctions based on topological insulators (TIs). 
The surface states of a three-dimensional TI give rise to the formation of  topologically nontrivial helical Andreev bound states with energy $\varepsilon=\pm\Delta\cos\phi/2$~\cite{fu_superconducting_2008}, i.e., they exhibit a zero-energy crossing at $\phi=\pi$. In contrast to the S-N-S case where even weak backscattering leads to a splitting of the accidental degeneracy of Andreev bound states at $\phi=\pi$, the crossing in the S-TI-S case is robust due to topological protection.
It gives rise to a $4\pi$-periodic Josephson current~\cite{fu_josephson_2009}. The latter is difficult to observe in experiment~\cite{maier_induced_2012, oostinga_josephson_2013} as it is a single-channel effect and subject to quasiparticle poisoning. 
Although recent experiments provide some evidence for the existence of helical Andreev bound states in the non-sinusoidal current phase relation \cite{sochnikov_nonsinusoidal_2015}, in the diffraction pattern~\cite{kurter_evidence_2015} and missing Shapiro steps in the ac Josephson effect~\cite{wiedenmann_4-periodic_2016}, it is still an outstanding challenge to distinguish between topologically trivial and nontrivial $4\pi$ modes. 

\begin{figure}

	\includegraphics[width=.46\textwidth]{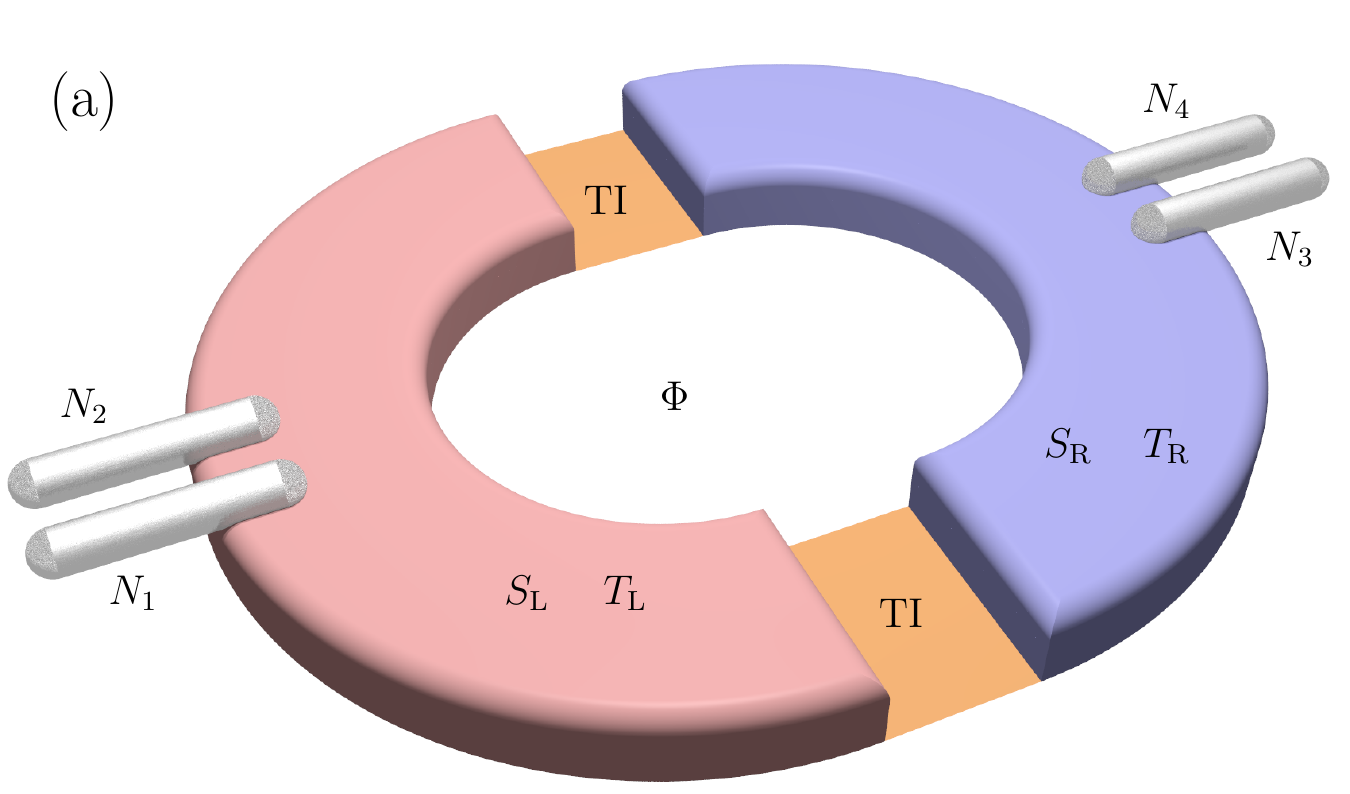}

	\includegraphics[width=.4\textwidth]{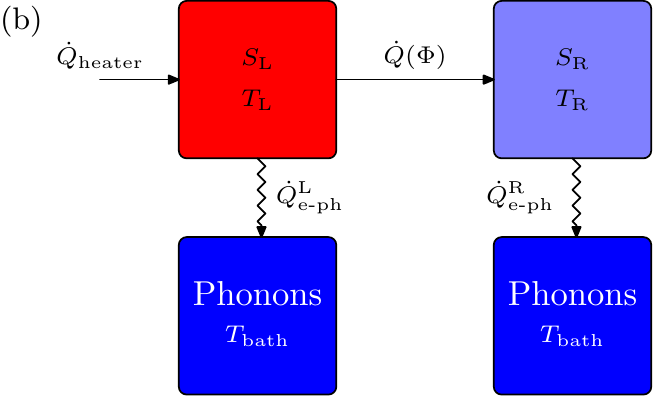}

	\caption{\label{fig:experiment}(a) SQUID geometry to detect phase-dependent heat currents. A magnetic flux $\Phi$ controls the phase difference across two identical Josephson junctions connecting superconductors $S_\text{L}$ and $S_\text{R}$ at temperatures $T_\text{L}$ and $T_\text{R}$, respectively. Normal metal contacts $N_{1-4}$ serve as heater and thermometer for the left and right superconductor, respectively. (b) Thermal model accounting for heat flows, see text for details.}

\end{figure}

Interestingly, not only the Josephson current but also the heat current between two superconductors kept at different temperatures depends on the phase difference across the junction. The effect arises due to the interference between quasiparticles carrying heat and Cooper pairs carrying phase information. It was theoretically predicted for tunnel junctions~\cite{maki_entropy_1965,maki_entropy_1966,guttman_phase-dependent_1997,guttman_interference_1998,golubev_heat_2013} as well as for point contacts with arbitrary transmission~\cite{zhao_phase_2003,zhao_heat_2004}. 
Phase-dependent heat currents can be used as an alternative approach to probe Andreev bound states. 
Indeed their effects have been very recently observed experimentally in a superconducting quantum interference device (SQUID) based on S-N-S junctions~\cite{giazotto_josephson_2012}. 
Subsequent experiments demonstrated the diffraction of heat currents in large junctions subject to magnetic fields~\cite{martinez-perez_quantum_2014} and realized a double SQUID that allows for enhanced control of heat flows~\cite{fornieri_nanoscale_2015}.
Theoretically, phase-dependent heat transport has been investigated in ferromagnetic Josephson junctions~\cite{giazotto_phase-tunable_2013,bergeret_phase-dependent_2013} as well as for ac-driven systems~\cite{virtanen_thermal_2014}. Furthermore, the fluctuations of phase-dependent heat currents~\cite{virtanen_fluctuation_2015} and their influence on the dephasing of flux qubits~\cite{spilla_measurement_2014} have been studied.

In this Rapid Communication, we demonstrate that phase-dependent heat currents provide a robust tool to probe the existence of topological Andreev bound states in S-TI-S Josephson junctions as well as to distinguish them from trivial $4\pi$ periodic Andreev bound states.
Since heat currents are carried by quasiparticles with energies above or below the superconducting gap, they provide complementary information to the Josephson current which is due to the formation of Andreev bound states within the gap.
For the same reason, heat transport does not suffer from quasiparticle poisoning in contrast to the $4\pi$-periodic Josephson effect.
We find that the thermal conductance carries clear signatures of the helical Andreev bound states both for short and long junctions as well as in the one-dimensional and two-dimensional cases.
Finally, we propose an experimental setup based on a SQUID, cf. Fig.~\ref{fig:experiment}, that allows for the detection of phase-dependent heat currents within the reach of state-of-the-art experimental technology.

\paragraph{Model.--}
We consider a ballistic Josephson junction \footnote{The diffusive regime is detrimental to the $p$-wave component of superconductivity that yields the topological Andreev bound states, see for example Ref.~\cite{tkachov_suppression_2013}.} based on the surface states of a three-dimensional TI in the $x-y$ plane. The areas $|x|>L/2$ are covered by conventional BCS superconductors with pair potential $\Delta e^{i\phi_r}$ where $\phi_r$ is the phase of superconductor $r=\text{L,R}$ such that there is a phase difference $\phi=\phi_\text{R}-\phi_\text{L}$ across the junction. For $|x|<L/2$, the pair potential vanishes, $\Delta=0$.
We assume that the pair potential changes on a length scale much shorter than the superconducting coherence length. This allows us to approximate the pair potential as a step function at the S-TI interface. We, furthermore, neglect proximity effects that would require a self-consistent determination of the pair potential.

Electron- and hole-like quasiparticles are described by the Bogoliubov-de Gennes Hamiltonian~\cite{blonder_transition_1982}
\begin{equation}
	H=\left(
	\begin{array}{cc}
		\hat h_{\vec k} & i\hat\sigma_y\Delta e^{i\phi_r}\\
		-i\hat\sigma_y\Delta e^{-i\phi_r} & -\hat h_{-\vec k}^*
	\end{array}
	\right),
\end{equation}
acting on wave functions describing electron- and hole-like quasiparticles with spin $\up$ and $\down$.
The single-particle Dirac Hamiltonian $\hat h_{\vec k}=\hbar v_\text{F}\vec k\cdot\hat{\boldsymbol\sigma}-\mu\hat\sigma_0$ describes the helical surface states of the TI in the absence of superconductivity where $v_\text{F}$ is the Fermi velocity, $\vec k$ the charge carrier wave vector, $\hat{\boldsymbol\sigma}$ the vector of Pauli matrices in spin space with $\sigma_0$ denoting the unit matrix and $\mu$ the chemical potential.

The eigenfunctions of the Bogoliubov-de Gennes Hamiltonian describing right-moving electron- and left-moving hole-like quasiparticles of energy $\omega$ are given by
$\psi_1(x,y)=(u,e^{i\theta_e}u,-e^{-i\phi_r}e^{i\theta_e}v,e^{-i\phi_r}v)^Te^{i\vec k_e\cdot\vec r}$ and 
$\psi_2(x,y)=(v,e^{i\theta_h}v,-e^{-i\phi_r}e^{i\theta_h}u,e^{-i\phi_r}u)^Te^{i\vec k_h\cdot \vec r}$, respectively, where $\vec r=(x,y)$, $\vec k_{e,h}=k_{e,h}(\cos\theta_{e,h},\sin\theta_{e,h})$, and $u=\frac{1}{2}\sqrt{1+\frac{\sqrt{\omega^2-\Delta^2}}{\omega}}$ and $v=\frac{1}{2}\sqrt{1-\frac{\sqrt{\omega^2-\Delta^2}}{\omega}}$ denote the usual coherence factors. The eigenfunctions $\psi_3(x,y)$ and $\psi_4(x,y)$ describing quasiparticles moving in the opposite direction are obtained from $\psi_1(x,y)$ and $\psi_2(x,y)$ by replacing the angle of incidence $\theta_{e,h}$ with $\pi-\theta_{e,h}$.
In the following, we will assume that the superconducting regions are heavily doped, $\mu\gg\Delta,\omega$, such that we can approximate $k_e=k_h\equiv k_\text{S}$ and $\theta_e=\theta_h\equiv\theta_\text{S}$.

Let us now consider the situation of an electron-like quasiparticle incident from the left-hand side. It gives rise to reflected and transmitted electron- and hole-like quasiparticles such that the wave function in the two superconductors reads $\psi_\text{L}(x,y)=\psi_1(x,y)+r_e\psi_3(x,y)+r_h\psi_2(x,y)$ and 
$\psi_\text{R}(x,y)=t_e\psi_1(x,y)+t_h\psi_4(x,y)$, respectively.
For a short junction, $L=0$, we model the interface barrier by the delta potential, $U\delta(x)$, with barrier height $U$, although our results are qualitatively independent of the barrier shape. The potential leads to the boundary condition $\psi_\text{L}(0,y)=(\cos Z \hat\tau_0\hat\sigma_0+i\sin Z\hat\tau_z\hat\sigma_x)\psi_\text{R}(0,y)$ where $Z=U/\hbar v_\text{F}$ and $\hat{\boldsymbol\tau}$ denotes the vector of Pauli matrices in particle-hole space. 
In a long junction, scattering arises due to the wave vector mismatch between the normal and superconducting regions. For simplicity, we assume that there is no  additional potential barrier at the interface. Thus, in this case, wave functions are continuous at the interface.
From the boundary conditions, the transmission amplitudes and subsequently the transmission probability for electron- and hole-like quasiparticles $\mathcal T_{e,h}(\omega,\phi)$ can be determined.

The thermal conductance of a one-dimensional Josephson junction is given by
\begin{equation}
	\kappa(\phi)=\frac{1}{h}\int_\Delta^\infty d\omega\; \omega \left[\mathcal T_e(\omega,\phi)+\mathcal T_h(\omega,\phi)\right] \frac{df}{dT},
\end{equation}
where $f=[\exp(\omega/\kBT)+1]^{-1}$ denotes the equilibrium Fermi distribution. For a two-dimensional Josephson junction, one has to perform an additional average over $\sin\theta_\text{S}$ and multiply with the number $N\gg1$ of open transport channels.

\paragraph{Short junction.--}
\begin{figure*}
	\includegraphics[width=.49\textwidth]{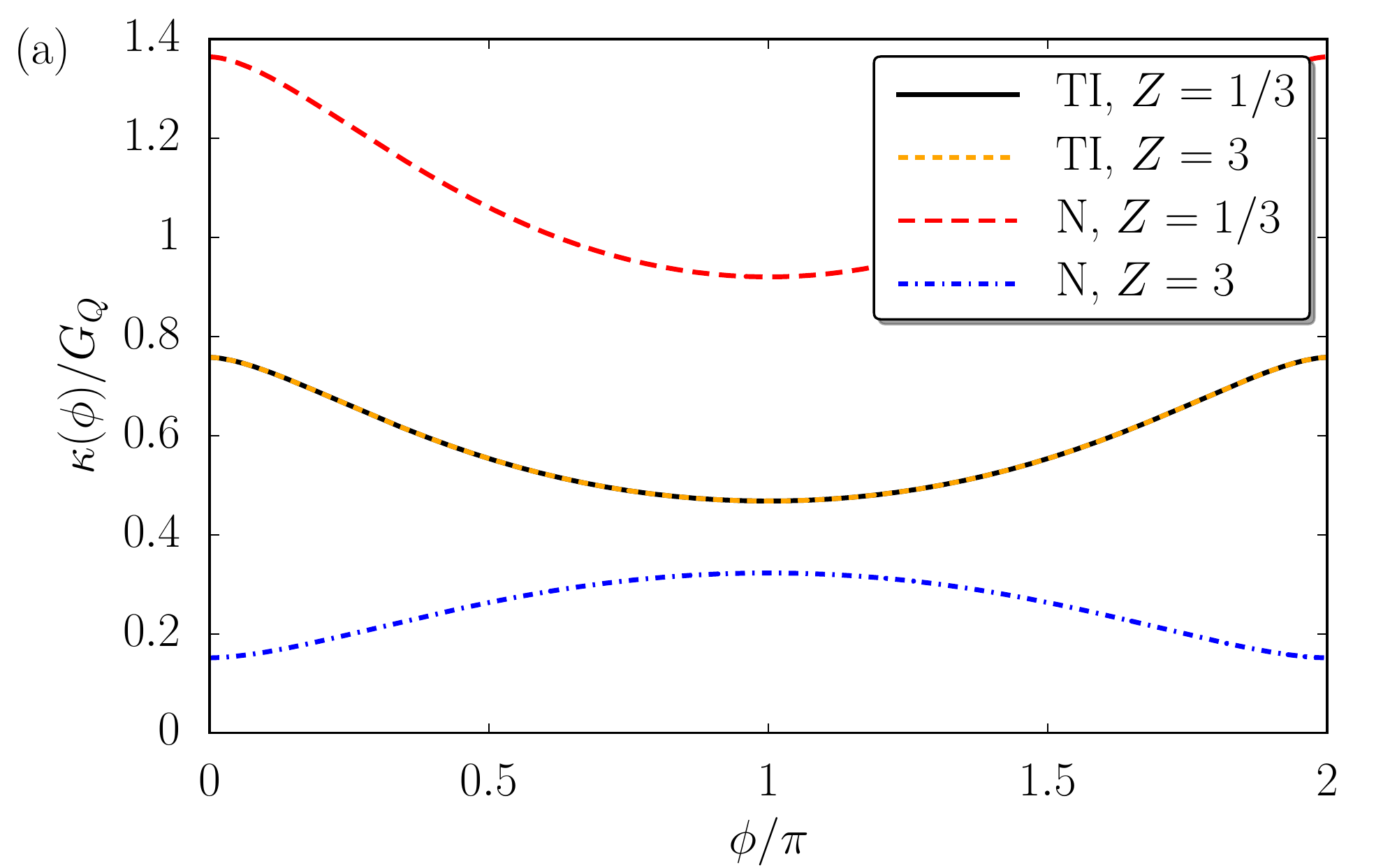}
	\includegraphics[width=.49\textwidth]{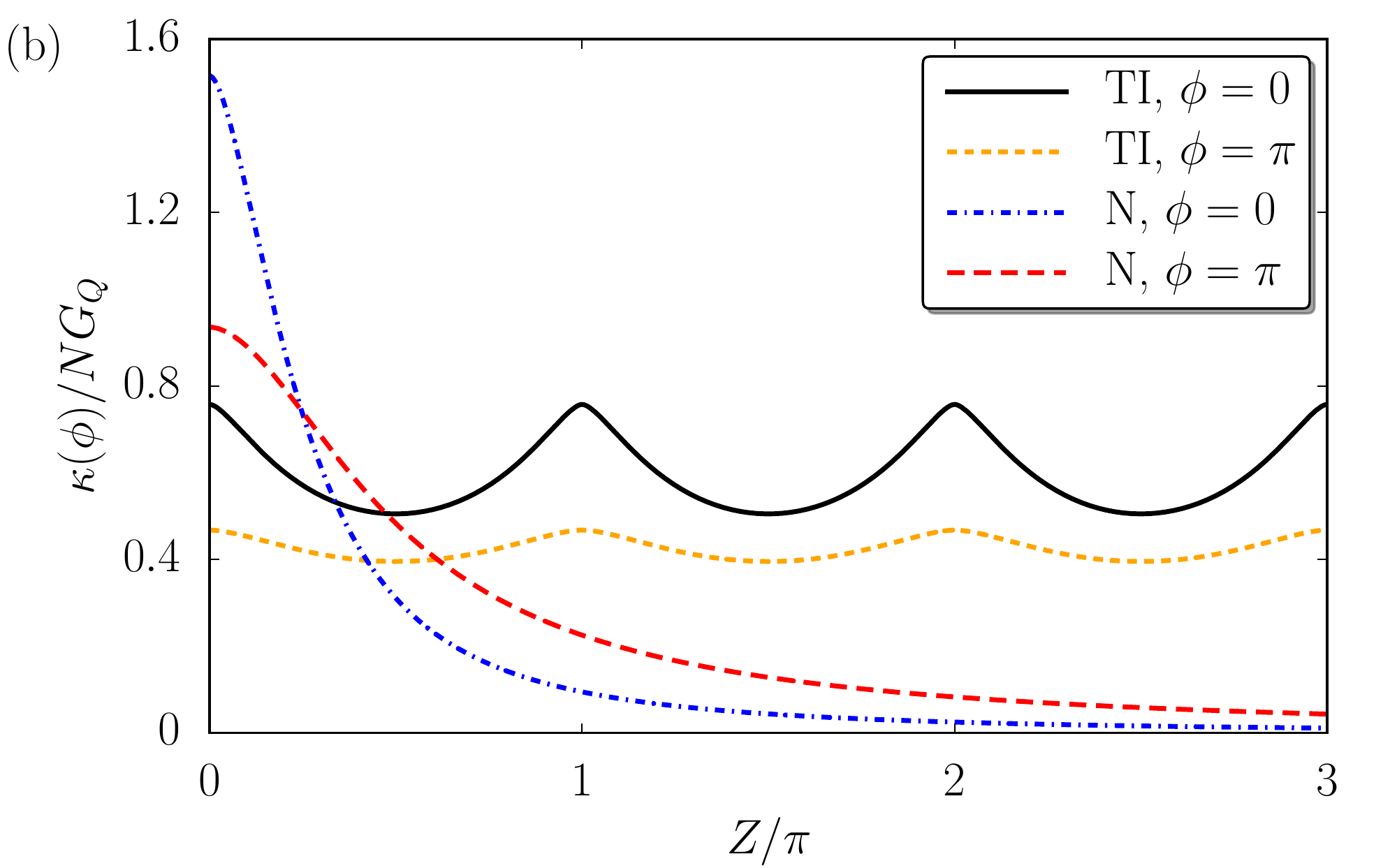}
	\caption{\label{fig:short}(a) Phase-dependence of the thermal conductance in units of the thermal conductance quantum $G_Q=\pi^2\kB^2T/(3h)$ for short, one-dimensional S-N-S and S-TI-S Josephson junctions for different interface barrier strength $Z$. (b) Thermal conductance for short, two-dimensional S-N-S and S-TI-S Josephson junctions with $N\gg1$ channels as a function of the interface barrier strength.  The temperature is chosen as $\kBT=\Delta/2$.}
\end{figure*}
For a short, one-dimensional S-TI-S junction, the transmission probability of quasiparticles is given by
\begin{equation}
	\mathcal T_{e,h}(\omega,\phi)=\frac{\omega^2-\Delta^2}{\omega^2-\Delta^2\cos^2\frac{\phi}{2}}.
\end{equation}
It is independent of the potential barrier at the interface, similarly to superconducting Klein tunneling for Andreev bound states~\cite{tkachov_helical_2013} and does not depend on the detailed spatial profile of the superconducting gap~\cite{sup_mat}.
In consequence, the thermal conductance of the junction exhibits a universal behavior independent of the interface scattering, cf. Fig.~\ref{fig:short}(a). In the limit $x=\Delta/\kBT\gg1$, we obtain the analytical approximation
\begin{align}
	\kappa(\phi=0)&=\frac{2\kB^2T}{h}(x^2+2x+2)e^{-x},\\
	\kappa(\phi=\pi)&=\frac{4\kB^2T}{h}(x+1)e^{-x},
\end{align}
i.e., lowering the temperature increases the amplitude of the thermal conductance oscillations but also reduces the total heat flow due to the reduced number of thermally excited quasiparticles~\cite{sup_mat}.

The phase-dependence of the S-TI-S junction  is in clear contrast to the S-N-S case. The latter exhibits a transition from a minimal thermal conductance at $\phi=\pi$ for a transparent interface to a maximal thermal conductance at $\phi=\pi$ in the tunneling limit~\cite{zhao_phase_2003,zhao_heat_2004}, cf. Fig.~\ref{fig:short}(a).
The different behaviors can be understood by analyzing the density of states at the interface. For an S-TI-S junction, the density of states is given by
\begin{equation}
	\rho(\omega)=\rho_\text{N}\mathcal T_{e,h}(\omega,\phi)\frac{|\omega|}{\sqrt{\omega^2-\Delta^2}},
\end{equation}
where $\rho_\text{N}$ is the interface density of states in the normal state.
At $\phi=0$, we recover the usual BCS density of states with a singularity at $\omega=\Delta$ due to the topological Andreev bound states merging with the continuum, similar to the S-N-S junction.
For transparent S-N-S and S-TI-S junctions the low-energy Andreev bound states remove density of states from above the gap  leading to a reduced thermal conductance.
However, in the tunneling regime S-TI-S junctions show strikingly different behavior from S-N-S junctions. This is because helical Andreev bound states are protected  by time-reversal symmetry against backscattering independent of the height of the barrier \cite{tkachov_helical_2013, tkachov_spin-helical_2013}.  
Therefore, they always stay in the middle of the superconducting gap leading to the suppression of the density of states 
and the minimum in thermal conductance for $\phi=\pi$. This is in clear contrast to the S-N-S junction, where the Andreev bound states shift to the edge of the superconducting gap leading to an enhancement of the thermal conductance at $\phi=\pi$~\cite{zhao_phase_2003,zhao_heat_2004}.

\begin{figure*}
	\includegraphics[width=.49\textwidth]{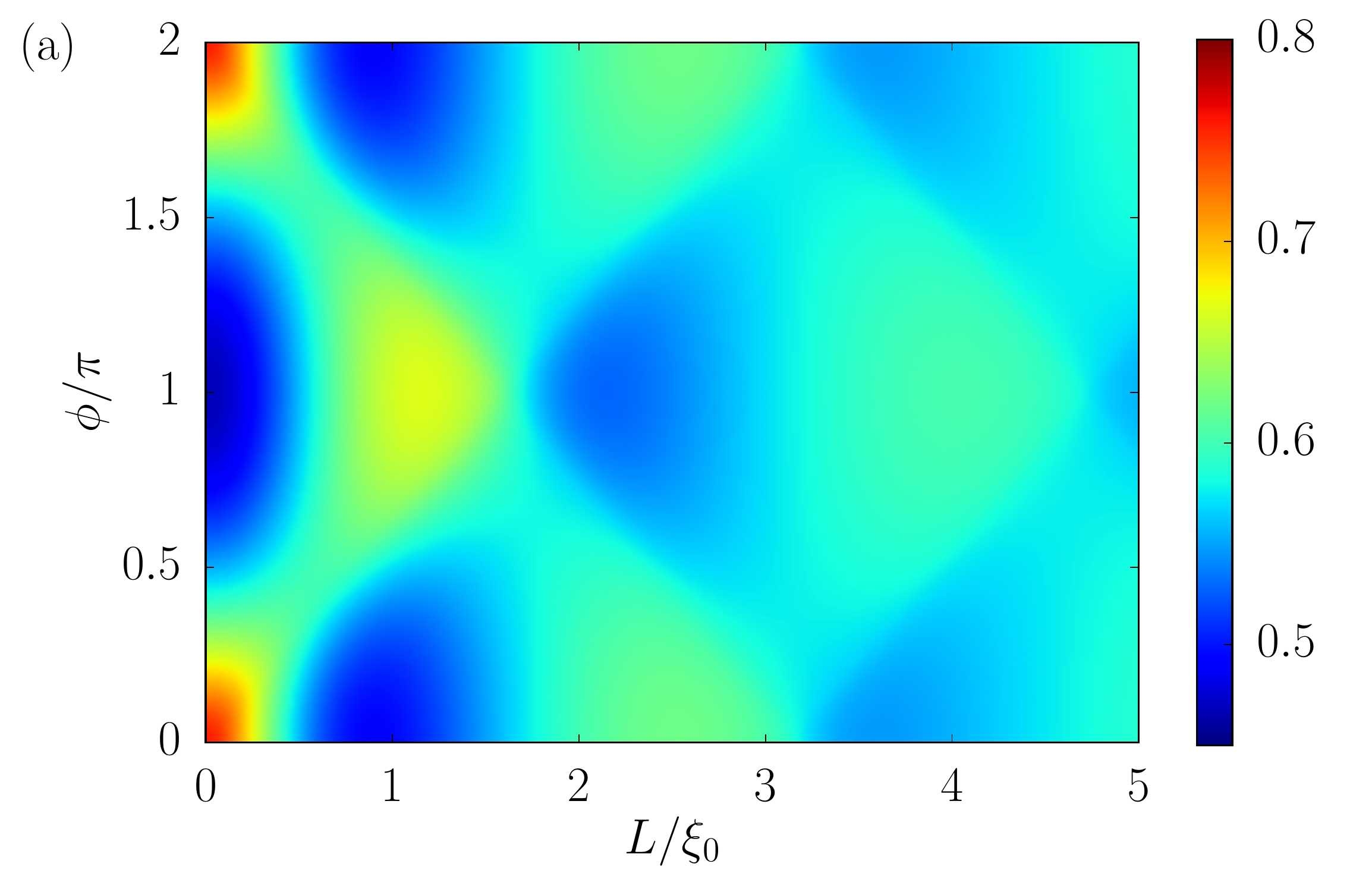}
	\includegraphics[width=.49\textwidth]{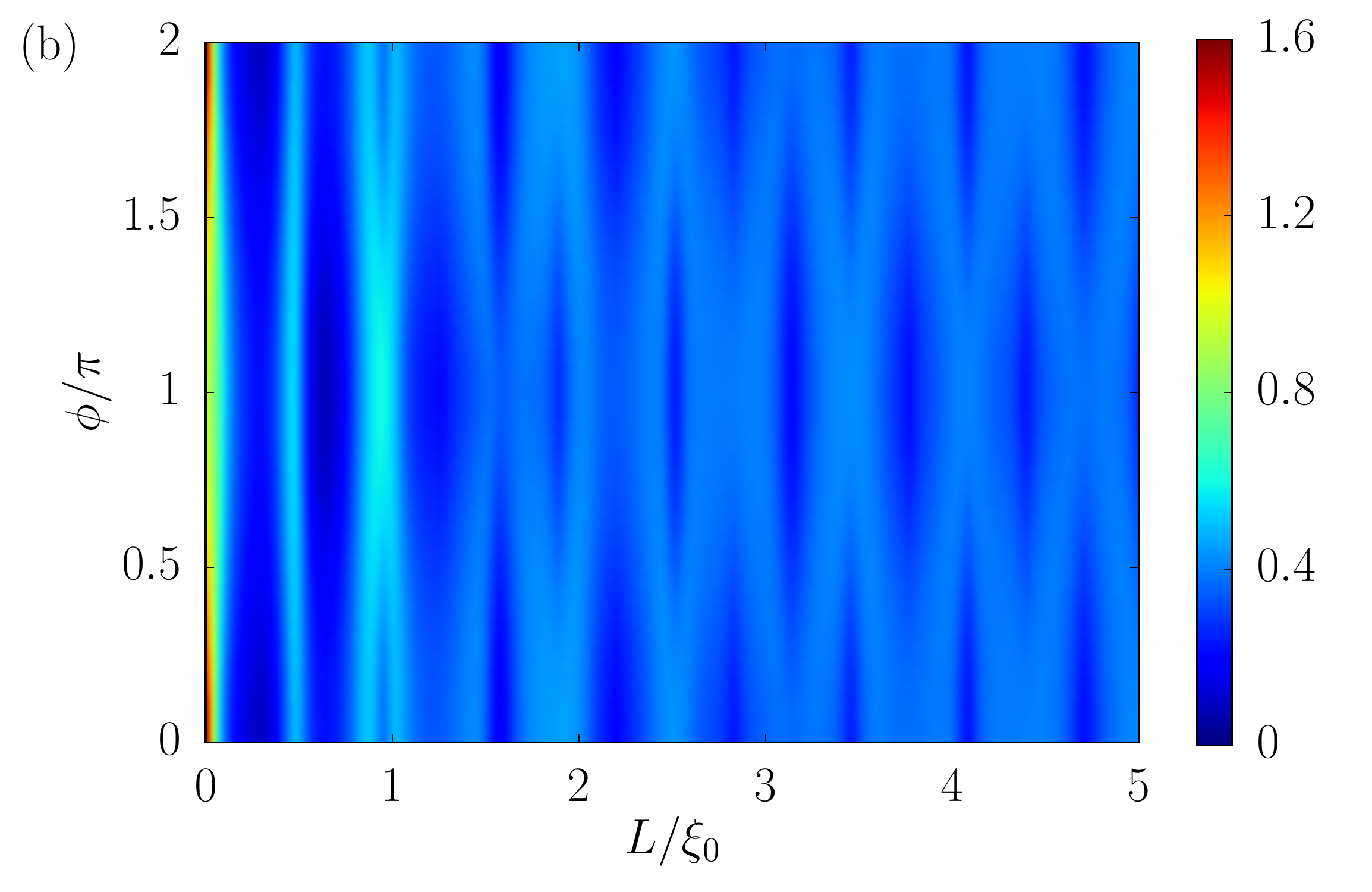}
	\caption{\label{fig:long}(a) Thermal conductance of a long, one-dimensional S-TI-S Josephson junction in units of $G_Q$ as a function of junction length $L$ and phase difference $\phi$ for $\kBT=\Delta/2$. (b) Thermal conductance of a long, one-dimensional S-N-S junction in units of $G_Q$ with $k_\text{N}=0.1k_\text{S}$, $\hbar v_\text{F}k_\text{S}=50\Delta$ and $\kBT=\Delta/2$.}
\end{figure*}
Due to their simplicity, one-dimensional systems allow for a transparent discussion of the underlying physics. 
At the same time, experiments on S-TI-S junctions typically realize two-dimensional setups due to issues with confinement in one dimension.
A two-dimensional S-N-S junction behaves qualitatively similar to its one-dimensional counterpart. It exhibits a transition from a minimal to a maximal thermal conductance at $\phi=\pi$ with increasing strength of the interface potential barrier, cf. Fig.~\ref{fig:short}(b). 
Since quasiparticles with oblique incidence experience an effectively higher potential barrier at the interface, the crossover occurs at smaller values of $Z$ in the two-dimensional case.
In a two-dimensional S-TI-S junction, only quasiparticles with normal incidence exhibit unit transmission while quasiparticles with oblique incidence can be backscattered. 
Nevertheless, the S-TI-S junction \emph{always} exhibits a minimal thermal conductance at $\phi=\pi$, in contrast to an S-N-S junction, see Fig.~\ref{fig:short}(b).
It originates from the fact that the heat current contribution for oblique incidence is geometrically suppressed compared to that for normal incidence. This leads to a dominant contribution of the topological Andreev bound state even in the two-dimensional case.
This remarkable finding  allows one to distinguish between topological and trivial $4\pi$ modes in two-dimensional junctions by tuning the junction transparency via a gate voltage acting on the central part of the junction. 
Further, the thermal conductance of an S-TI-S junction oscillates as a function of the barrier strength due to the formation of resonances at the interface. This is a unique feature of the scattering of Dirac quasiparticles.
To summarize, we find that for short S-TI-S junctions there are clear signatures of helical Andreev bound states for both the one- and the two-dimensional case.

\paragraph{Long junction.--}
For a long, one-dimensional S-TI-S junction, the energy dependence of the electron and hole wave vector in the intermediate TI region, $k_\text{N,e/h}=k_\text{N}\pm\omega/(\hbar v_\text{F})$, where $k_\text{N}$ is the Fermi wave vector, gives rise to a modified transmission function,
\begin{equation}
	\mathcal T_{e,h}(\omega,\phi)=\frac{\omega^2-\Delta^2}{\omega^2-\Delta^2\cos^2\left(\frac{\phi}{2}\mp\frac{\omega L}{\hbar v_\text{F}}\right)}.
\end{equation}
As a consequence, the maximal thermal conductance no longer occurs at $\phi=0$ but is shifted to finite values of $\phi$, cf. Fig.~\ref{fig:long}(a). This is reminiscent of the $\phi$-junction behavior of the Josephson current in certain Josephson junctions with singlet and triplet pairing~\cite{buzdin_direct_2008,brunetti_anomalous_2013,yokoyama_anomalous_2014,feinberg_spontaneous_2014,buzdin_periodic_2003,pugach_method_2010,goldobin_josephson_2011,sickinger_experimental_2012,kulagina_spin_2014,sothmann_josephson_2015}. We remark that the $\phi$-junction behavior of the thermal conductance here is due to the wave vector difference between electrons and holes and, thus, of a completely different origin than the $\phi$-junction behavior of the Josephson current in the aforementioned junctions.
Furthermore, the oscillations of the thermal conductance with $\phi$ are damped on the scale of the superconducting coherence length $\xi_0=\Delta/(\hbar v_\text{F})$.

For an S-N-S junction, the energy dependence of electron and hole wave vectors in the normal region of the junction also gives rise to a $\phi$-junction behavior of the heat conductance just as for an S-TI-S junction.  On top of this, the S-N-S junction exhibits Fabry-Pérot type oscillations that occur on a length scale $k_\text{N}^{-1}$ due to finite scattering probabilities at the S-N interfaces. The interplay between these two effects gives rise to a complicated interference pattern, cf. Fig.~\ref{fig:long}(b).
Hence, a distinct $\phi$-junction behavior of the thermal conductance provides a clear signature of a topological Josephson junction.

\paragraph{Possible experiment.--}
In order to experimentally confirm our theoretical predictions, we suggest using a SQUID consisting of two superconducting electrodes connected via identical S-TI-S Josephson junctions, cf. Fig.~\ref{fig:experiment}(a).
The phase difference across the junctions can be controlled by a magnetic flux $\Phi$, while the junction transparency can be tuned via gate voltages~\cite{hart_controlled_2015}.
Control over the junction length can be achieved by growing samples with different $L$ on the same wafer such that other junction properties are very similar.
The heat flow can be accounted for in a simple thermal model, see Fig.~\ref{fig:experiment}(b), with the heat $\dot Q_\text{heater}$ injected by a heater, the phase-dependent heat flow $\dot Q(\Phi)=2\kappa(\phi)(T_\text{L}-T_\text{R})$ through the two Josephson junctions as well as heat losses into the substrate due to electron-phonon coupling, $\dot Q^r_\text{e-ph}=0.98 e^{-\Delta/\kBT_r}\Sigma_r V_r(T_r^5-T_\text{bath}^5)$, $r=\text{L,R}$~\cite{timofeev_recombination-limited_2009}.
In the following, we assume an electron-phonon coupling strength of $\Sigma_r=\unit[10^9]{Wm^{-3}K^{-5}}$ and take the volume of each superconductor to be $V_r=\unit[10^{-20}]{m^3}$~\cite{giazotto_josephson_2012}.
For a bath temperature of $T_\text{bath}=\unit[100]{mK}$, a temperature of the left superconductor $T_\text{L}=\unit[500]{mK}$ and a superconducting pair potential equal to twice the average electron temperature, we find that the temperature of the right superconductor varies between \unit[360]{mK} and \unit[380]{mK} for single-channel Josephson junctions.
This variation is clearly within the reach of current experimental sensitivity which is around \unit[100]{$\mu$K}~\cite{giazotto_josephson_2012} such that our predicted effect is observable.
We remark that while the precise value of the temperature of the right superconductor depends on the exact values of parameters, the variation of about \unit[20]{mK} is very robust with respect to parameter variations and occurs, e.g., also for junctions with many open transport channels.

\paragraph{Summary.--}
We demonstrated that the phase-dependent thermal conductance of an S-TI-S Josephson junction contains clear signatures of the existence of topological Andreev bound states. Importantly, the predicted effects are robust with respect to the dimensionality and length of the junction. 
Furthermore, we have proposed an experimental setup that permits one to verify of our predictions using available SQUID devices. Our results allow one to observe experimentally a clear difference between topologically trivial and nontrivial $4\pi$ modes and open the perspective of studying exotic superconductivity in various materials via heat transport.

\acknowledgments
\paragraph{Acknowledgments.--}
We thank F. Giazotto and L. W. Molenkamp for interesting discussions and acknowledge financial support from the DFG via SFB 1170 "ToCoTronics", and the ENB Graduate School on Topological Insulators.

% \bibliographystyle{apsrev4-1}
% \bibliography{/home/bjoern/LaTeX/Bibtex/Meine_Bibliothek}


\newpage


\section*{Supplementary material}

# Supplemental Material for:
# Fingerprint of topological Andreev bound states in phase-dependent heat transport

Björn Sothmann and Ewelina M. Hankiewicz

*Institute for Theoretical Physics and Astrophysics, University of Würzburg, Am Hubland, 97074 Würzburg, Germany*

(Dated: July 23, 2016)

## I. TEMPERATURE DEPENDENCE OF THE THERMAL CONDUCTANCE

In general, there is no closed analytical expression for the thermal conductance of the short S-TI-S Josephson junction. However, in the limit of a short, one-dimensional junction, we find that for $\phi = 0$ and $\phi = \pi$ the thermal conductance is given by

$$\kappa(\phi = 0) = \frac{2k_\mathrm{B}^2 T}{h} \left\{ x \left[ \frac{x}{1+e^x} + 2\log\left(1+e^{-x}\right) \right] - 2\,\mathrm{Li}_2\left(-e^{-x}\right) \right\}, \tag{1}$$

$$\kappa(\phi = \pi) = \frac{4k_\mathrm{B}^2 T}{h} \left[ x \log\left(1+e^{-x}\right) - \mathrm{Li}_2\left(-e^{-x}\right) \right], \tag{2}$$

where $x = \Delta/k_\mathrm{B} T$ and $\mathrm{Li}_2(z) = \sum_{k=1}^\infty \frac{z^k}{k^2}$ denotes the dilogarithm. In the limit of low temperatures, $x \gg 1$, these expressions simplify to the result given in the main text,

$$\kappa(\phi = 0) \approx \frac{2k_\mathrm{B}^2 T}{h}(x^2 + 2x + 2)e^{-x}, \tag{3}$$

$$\kappa(\phi = \pi) \approx \frac{4k_\mathrm{B}^2 T}{h}(x+1)e^{-x}. \tag{4}$$

The temperature dependence of the thermal conductance of a short, one-dimensional junction is shown in Fig. 1. As the temperature is lowered, the oscilllations with $\phi$ are increased. At the same time, the thermal conductance is strongly reduced to the number of thermally excited quasiparticles being exponentially suppressed.

For a two-dimensional junction, there is no analytical expression for the thermal conductance any longer. However, numerically we find that a qualitatively similar behavior as in the one-dimensional case emerges, i.e, while the oscillations become more pronounced upon reducing the temperature, the thermal conductance is exponentially suppressed, see Fig. 2.

## II. SPATIAL VARIATION OF THE ORDER PARAMETER

In principle, one can calculate the superconducting order parameter that enters the Bogoliubov-de Gennes equation in a self-consistent fashion. However, this approach does not allow for an analytical solution and instead relies entirely on numerical calculations. The latter have shown that the superconducting gap decreases slightly from its bulk value on a length scale given by the superconducting coherence length as one approaches the interface. Right at the interface, there is a sharp decrease of the gap on a length scale given Fermi wave length [1]. Taking into account a realistic spatial dependence of the superconducting gap is still challenging to treat analytically within the Bogoliubov de Gennes approach. To this end, here we consider instead a simple toy model consisting of a short, one-dimensional S-S'-TI-S'-S junction. Here, S denotes a superconductor with the constant bulk gap $\Delta$. The second superconductor S' (of arbitrary length) has a smaller gap $\Delta'$ and thus represents the region with a reduced gap near the interface. Finally, we model the central nonsuperconducting region

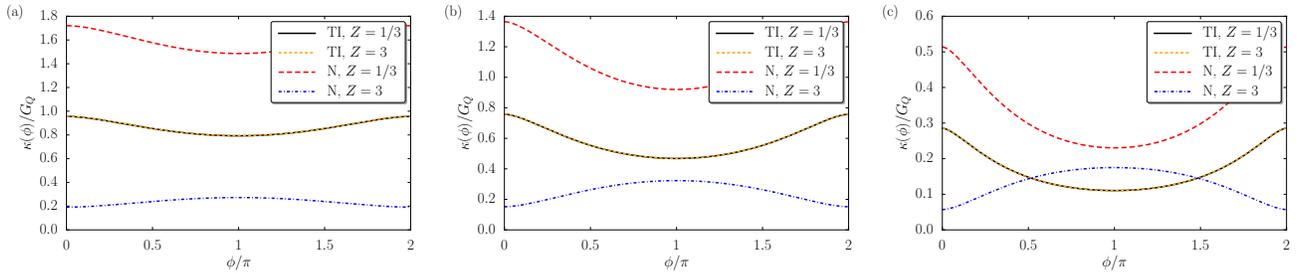

FIG. 1. Phase-dependent thermal conductance for short, one-dimensional S-N-S and S-TI-S Josephson junctions in units of the thermal conductance quantum $G_Q$ for (a) $k_\mathrm{B}T = \Delta$, (b) $k_\mathrm{B}T = \Delta/2$ and (c) $k_\mathrm{B}T = \Delta/4$.

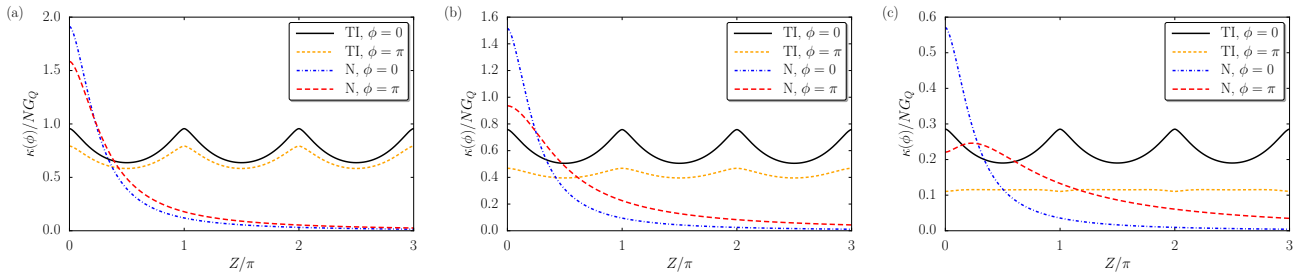

FIG. 2. Phase-dependent thermal conductance for short, two-dimensional S-N-S and S-TI-S Josephson junctions as a function of the interface barrier strength $Z$ in units of the thermal conductance quantum $G_Q$ for (a) $k_B T = \Delta$, (b) $k_B T = \Delta/2$ and (c) $k_B T = \Delta/4$.

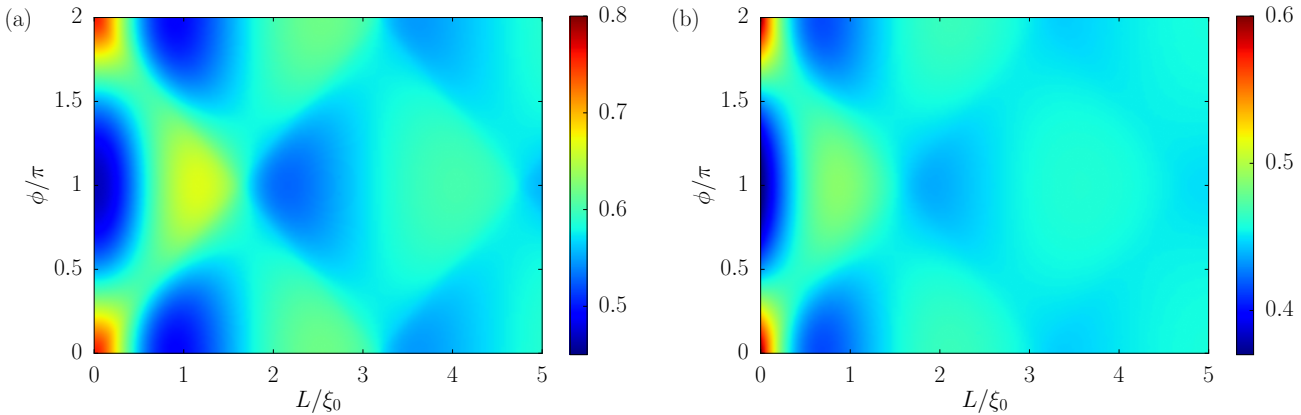

FIG. 3. Phase-dependent thermal conductance for a long (a) one-dimensional and (b) two-dimensional S-TI-S Josephson junctions with $N \gg 1$ channels in units of $N G_Q$ for $k_B T = \Delta/2$ and $k_N = k_S$.

via a delta potential just as we did for the simpler S-TI-S case in the main text. By matching wave functions at the different interfaces, we obtain for the transmission probability of an incoming quasiparticle

$$\mathcal{T}_{e,h}(\omega, \phi) = \frac{\omega^2 - \Delta^2}{\omega^2 - \Delta^2 \cos^2 \frac{\phi}{2}}, \tag{5}$$

i.e. we obtan the same result as for an S-TI-S junction. Importantly, the presence of the additional superconducting parts with a reduced gap does not influence the transmission probability. This is a consequence of superconducting Klein tunneling that ensures that incoming quasiparticles are transmitted with unit probability through the junction independet of the presence of scattering potentials inside the junction. Hence, we can conclude that a weak spatial variation of the order parameter has a negligible influence on the thermal transport through a topological Josephson junction.

### III. LONG TWO-DIMENSIONAL JUNCTION

In the main text, we restricted ourselves to the analysis of long, one-dimensional Josephson junctions. Here, we consider a two-dimensional S-TI-S Josephson junction of length L. We assume that the Fermi wave vector is the same in the superconducting and normal region of the junction, $k_N = k_S$. Under these conditions, we find that the transmission probability of a quasiparticle incident under an angle $\theta$ onto the interface is given by

$$\mathcal{T}_{e,h}(\omega) = \frac{\omega^2 - \Delta^2}{\omega^2 - \Delta^2 \cos^2 \left( \frac{\phi}{2} \mp \frac{\omega L}{\hbar v_F \cos \theta} \right)}. \tag{6}$$

The resulting thermal conductance is shown in Fig. 3 in direct comparison with the thermal conductance of a long, one-dimensional junction discussed in the main text. Importantly, we find that in both cases we obtain a qualitatively similar behavior, i.e., for finite length the maxima of the thermal conductance get shifted to finite values of $\phi$. Furthermore, we find that in both cases the oscillations are damped as the junction length increases. The only quantitative difference between the one-dimensional and two-dimensional case is that in the latter the damping of the oscillations is slightly stronger. The

physical picture behind this is that in a two-dimensional junction electrons impinging under a certain angle see a junction that is effectively longer. Hence, quasiparticles with different $\theta$ experience a different phase shift which gives rise to a smearing out of the oscillations when averaging over $\theta$.



% \includepdf{Supp_Mat_corr.pdf} 

\end{document}